\documentclass[showpacs,10pt,twocolumn,prb]{revtex4-1}

\usepackage{amsmath}
\usepackage{amssymb}
\usepackage{graphicx}
\usepackage{amssymb}
\usepackage{graphics}
\usepackage{epsfig}
\usepackage{CJK}
\usepackage{color}

\begin{document}

\begin{CJK*}{GBK}{Song}
\title{Anomalous Hall effect in trigonal Cr$_5$Te$_8$ single crystal}
\author{Yu Liu and C. Petrovic}
\affiliation{Condensed Matter Physics and Materials Science Department, Brookhaven National Laboratory, Upton, New York 11973, USA\\}
\date{\today}

\begin{abstract}
We report anomalous Hall effect (AHE) and transport properties of trigonal Cr$_5$Te$_8$ (tr-Cr$_5$Te$_8$) single crystals. The electrical resistivity as well as the Seebeck coefficient shows a clear kink at the paramagnetic-ferromagnetic transition of tr-Cr$_5$Te$_8$, which is also confirmed by the heat capacity measurement. The scaling behavior between anomalous Hall resistivity $\rho^A_{xy}$ and longitudinal resistivity $\rho_{xx}$ is linear below $T_c$. Further analysis suggests that the AHE in tr-Cr$_5$Te$_8$ is dominated by the skew-scattering mechanism rather than the intrinsic or extrinsic side-jump mechanism.
\end{abstract}

\maketitle
\end{CJK*}

\section{INTRODUCTION}

The anomalous Hall effect (AHE) is an important electronic transport phenomenon.\cite{Nagaosa} Compared with the ordinary Hall effect (OHE), originating from the deflection of charge carriers by the Lorentz force in a magnetic field, the AHE can arise because of two qualitatively different microscopic mechanisms: an intrinsic mechanism connected to the Berry curvature and extrinsic processes due to scattering effects.\cite{Nagaosa,Fang,Haldane,Xiao,Nakatsuji} Recently, the AHE in magnetic frustrated materials and/or noncollinear structure have attracted much attention, such as PdCrO$_2$ and Fe$_{1.3}$Sb with a triangular lattice,\cite{Takatsu,Shiomi} Pr$_2$Ir$_2$O$_7$ and Nd$_2$Mo$_2$O$_7$ with a pyrochlore lattice,\cite{Machida,Taguchi} Mn$_3$Sn and Mn$_3$Ge with a Kagome lattice,\cite{Kubler,Kiyohara,Nayak} and antiferromagnets with noncollinear spin structures.\cite{Ueland,Surgers,Suzuki}

Binary chromium tellerides Cr$_{1-x}$Te are ferromagnetic with $T_c$ of 170 $\sim$ 360 K depending on Cr occupancy.\cite{Herbert,Street,Hamasaki,Akram,Lukoschus,Huang1,Huang2} Cr$_{1-x}$Te with $x < 0.1$ crystallize in the hexagonal NiAs structure, while Cr$_3$Te$_4$ ($x = 0.25$) and Cr$_2$Te$_3$ ($x = 0.33$) form monoclinic and trigonal crystal structures where Cr vacancies occupy every second metal layer. Neutron-diffraction measurement shows that the saturation magnetization in Cr$_{1-x}$Te is small due to possible spin canting and itinerant nature of the $d$ electrons.\cite{Hamasaki,Andresen} Electron correlation effect in itinerant ferromagnets has also been discussed in the photoemission spectra.\cite{Shimada} For $x = 0.375$, the monoclinic phase (m-Cr$_5$Te$_8$) is stable in the range 59.6-61.5 atomic percent Te. A slight increase in Te content leads to an order-disorder transition from monoclinic to trigonal phase (tr-Cr$_5$Te$_8$). In tr-Cr$_5$Te$_8$ the Cr atoms are located on four crystallographically different sites leading to the formation of a five-layer superstructure of the CdI$_2$ type with $P\bar{3}m1$ space group [Fig. 1(a)]. There are triangular lattices formed by Cr atoms [Fig. 1(b)], suggesting geometric frustration in tr-Cr$_5$Te$_8$. The tr-Cr$_5$Te$_8$ shows a higher Curie temperature ($T_c \sim 237$ K) despite its lower Cr content.\cite{YuLIU} Their critical behavior and magnetocaloric properties are recently studied,\cite{XiaoZ,XiaoH} however, the transport properties are still unknown.

Here we investigate the AHE in tr-Cr$_5$Te$_8$ single crystal, in connection with its transport properties. The observed anomalies in $\rho(T)$ and $S(T)$ at $\sim$ 237 K reflects reconstruction of the Fermi surface, corresponding well to the paramagnetic-ferromagnetic (PM-FM) transition, which is also confirmed by $C_p(T)$. The linear dependence of the anomalous Hall resistivity $\rho^A_{xy}$ and the longitudinal resistivity $\rho_{xx}$ below $T_c$ indicates the skew-scattering mechanism dominates the AHE in tr-Cr$_5$Te$_8$.

\section{EXPERIMENTAL DETAILS}

Single crystals of tr-Cr$_5$Te$_8$ were fabricated by the self-flux method and characterized as described previously.\cite{YuLIU} The element ratio determined by x-ray energy-dispersive spectroscopy is Cr : Te = 0.62(3) : 1 [Fig. 1(c)], and it is referred to as tr-Cr$_5$Te$_8$ throughout this paper. The dc magnetization, electrical and thermal transport, and heat capacity were measured in the Quantum Design MPMS-XL5 and PPMS-9 systems. Single crystals were cut into rectangles with dimensions of 2 $\times$ 2.5 $\times$ 0.25 mm$^3$. The calculated demagnetization factor $N_d$ is about 0.8. Standard four-probe method was applied in the longitudinal and Hall resistivity measurement with in-plane current. In order to effectively eliminate the longitudinal resistivity contribution due to voltage probe misalignment, the Hall resistivity was calculated by the difference of transverse resistance measured at positive and negative fields, i.e., $\rho_{xy}(\mu_0H) = [\rho(+\mu_0H)-\rho(-\mu_0H)]/2$.

\section{RESULTS AND DISCUSSIONS}

\begin{figure}
\centerline{\includegraphics[scale=1]{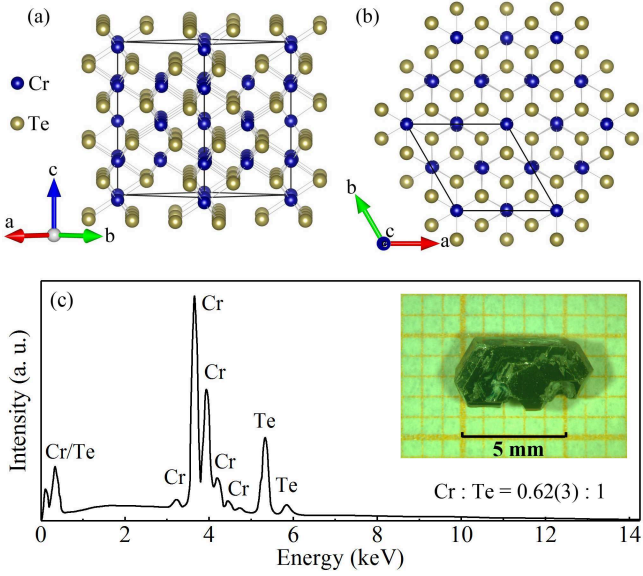}}
\caption{(Color online). Crystal structure of tr-Cr$_5$Te$_8$ from (a) side and (b) top view. (c) X-ray energy-dispersive spectroscopy of tr-Cr$_5$Te$_8$. Inset shows a photograph of tr-Cr$_5$Te$_8$ single crystal on a 1 mm grid.}
\label{structure}
\end{figure}

\begin{figure}
\centerline{\includegraphics[scale=1]{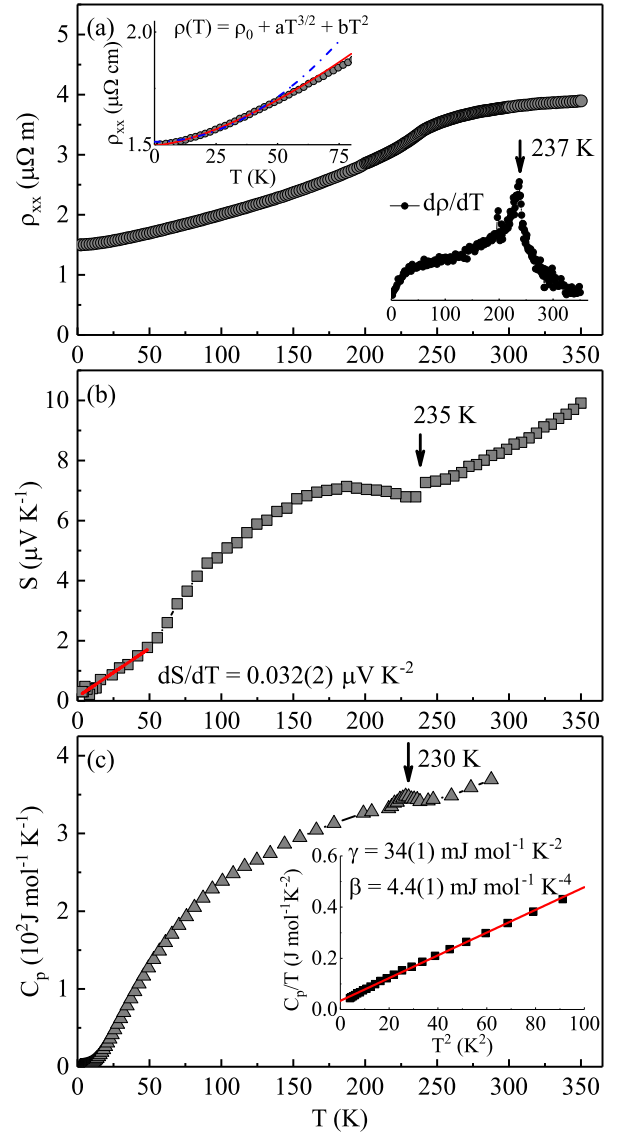}}
\caption{(Color online). Temperature dependence of (a) in-plane resistivity $\rho(T)$, (b) Seebeck coefficient $S(T)$, and (c) heat capacity $C_p(T)$ of tr-Cr$_5$Te$_8$ single crystal measured in zero field. Insets in (a) show the low temperature part fitted by $\rho(T) = \rho_0 + aT^{3/2} + bT^2$ (solid line), in comparison with $\rho(T) = \rho_0 + bT^2$ (dashed line), and the $d\rho/dT$ vs $T$ curve. Inset in (c) exhibits the low temperature $C_p(T)/T$ vs $T^2$ curve fitted by $C_p(T)/T = \gamma + \beta T^2$.}
\label{RSCT}
\end{figure}

Figure \ref{RSCT}(a) shows the temperature-dependent in-plane resistivity $\rho_{xx}(T)$ of tr-Cr$_5$Te$_8$, indicating a metallic behavior with a relatively low residual resistivity ratio [RRR = $\rho$(300 K)/$\rho$(2 K) = 2.5] most likely due to large Cr vacancies. A clear kink is observed at 237 K, which is determined by the maximum of the $d\rho/dT$ curve, corresponding well to the PM-FM transition. The renormalized spin fluctuation theory suggests that the electrical resistivity shows a $T^2$ dependence on the temperature $T$ for itinerant ferromagnetic system.\cite{Ueda1} In tr-Cr$_5$Te$_8$, the low temperature resistivity fitting gives a better result by adding an additional $T^{3/2}$ term,
\begin{equation}
\rho(T) = \rho_0 + aT^\frac{3}{2} + bT^2,
\end{equation}
where $\rho_0$ is the residual resistivity. The fitting yields $\rho_0$ = 1.50(1) $\mu\Omega$ cm, $a$ = 5.5(2)$\times$$10^{-4}$ $\mu\Omega$ cm K$^{-1}$, and $b$ = 1.0(8)$\times$10$^{-6}$ $\mu\Omega$ cm K$^{-2}$, indicating the $T^{3/2}$ term predominates. This means the interaction between conduction electrons and localized spins could not be simply treated as a small perturbation to a system of free electrons and strong electron correlation should be considered in tr-Cr$_5$Te$_8$.\cite{Liu}

The Seebeck coefficient $S(T)$ of tr-Cr$_5$Te$_8$ is positive in the whole temperature range, indicating dominant hole-type carriers [Fig. \ref{RSCT}(b)]. With temperature decrease, the value of $S(T)$ decreases gradually and shows a reduction at $T_c$, reflecting the reconstruction of the Fermi surface, and then changes slightly featuring a broad maximum around 180 K. Below 50 K, the diffusive Seebeck response of Fermi liquid dominates and is expected to be linear in $T$. In a metal with dominant single-band transport, the Seebeck coefficient could be described by the Mott relationship,
\begin{equation}
S = \frac{\pi^2}{3}\frac{k_B^2T}{e}\frac{N(\varepsilon_F)}{n},
\end{equation}
where $N(\varepsilon_F)$ is the density of states (DOS), $\varepsilon_F$ is the Fermi energy, $n$ is carrier concentration, $k_B$ is the Boltzman constant and $e$ is the absolute value of electronic charge.\cite{Barnard} The derived $dS/dT$ below 50 K is about 0.032(2) $\mu$V K$^{-2}$.

\begin{figure}
\centerline{\includegraphics[scale=1]{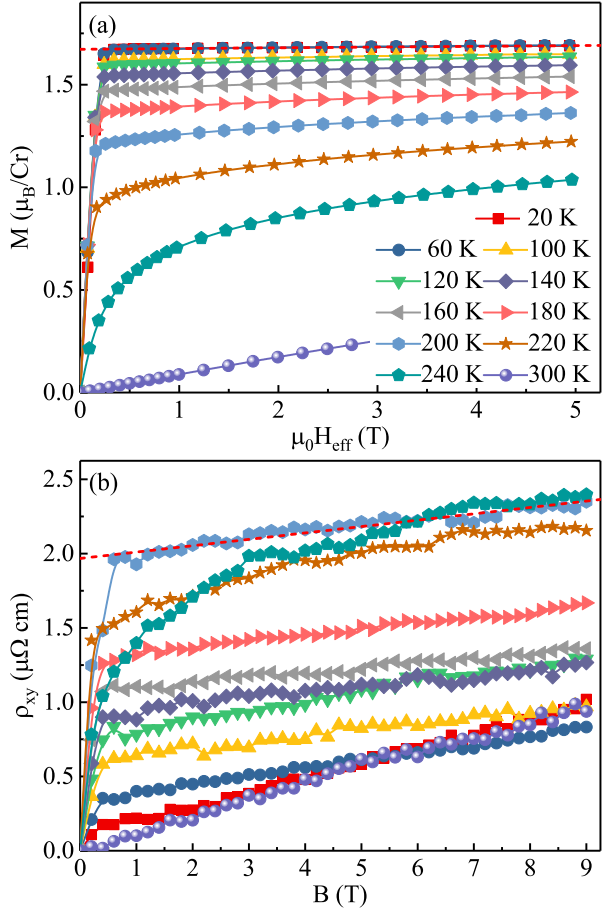}}
\caption{(Color online). (a) Effective field dependence of magnetization $M(\mu_0H_{eff})$ and (b) Hall resistivity $\rho_{xy}(B)$ as a function of magnetic induction $B$ for tr-Cr$_5$Te$_8$ single crystal at indicated temperatures with out-of-plane fields. The red dashed lines are linear fits of $M(\mu_0H_{eff})$ and $\rho_{xy}(B)$ at high field region.}
\label{RMH}
\end{figure}

\begin{figure}
\centerline{\includegraphics[scale=1]{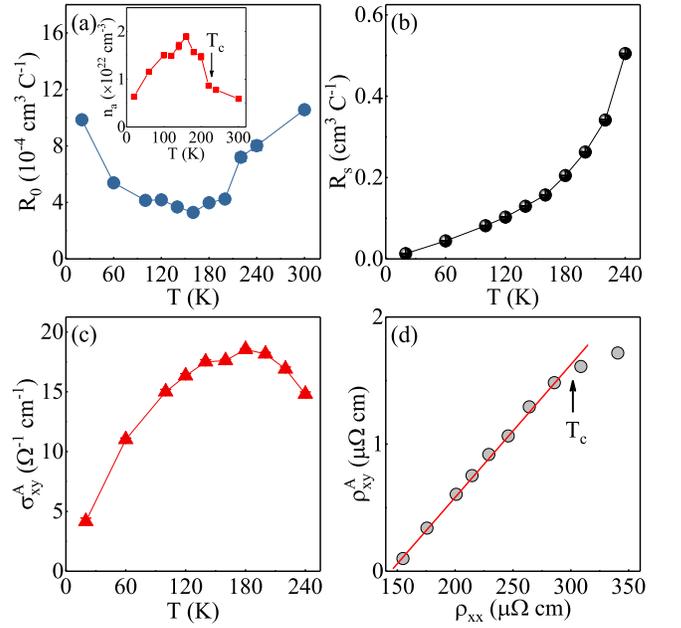}}
\caption{(Color online). Temperature dependence of (a) ordinary Hall coefficient $R_0(T)$ (left axis), derived carrier concentration $n_a(T)$ (right axis), and (b) anomalous Hall coefficient $R_s(T)$ fitted from $\rho_{xy}(B,T)$ using $\rho_{xy} = R_0B + R_s\mu_0M$. (c) Anomalous Hall conductivity $\sigma_{xy}^A$ (left axis) and scaling coefficient $S_H(T)$ (right axis) as a function of temperature. (d) Plot of $\rho^A_{xy}$ vs $\rho_{xx}$ with a linear fit (solid red line) below $T_c$.}
\label{parameters}
\end{figure}

Figure \ref{RSCT}(c) exhibits the temperature-dependent heat capacity $C_p(T)$ for tr-Cr$_5$Te$_8$, in which a clear peak was observed near the PM-FM transition. The high temperature data approach the Dulong Petit value of $3NR$ $\approx$ 324 J mol$^{-1}$ K$^{-1}$. The low temperature data from 2 to 10 K can be well fitted by $C_p/T = \gamma + \beta T^2$ [inset in Fig. 2(c)], where the first term is the Sommerfeld electronic specific heat coefficient and the second term is low-temperature limit of the lattice heat capacity. The obtained $\gamma$ and $\beta$ are 34(1) mJ mol$^{-1}$ K$^{-2}$ and 4.4(1) mJ mol$^{-1}$ K$^{-4}$, respectively. The Debye temperature $\Theta_D$ = 179(1) K can be derived from $\beta$ using $\Theta_D = (12\pi^4NR/5\beta)^{1/3}$, where $N$ is the number of atoms per formula unit and $R$ is the gas constant. The electronic specific heat
\begin{equation}
C_e=\frac{\pi^2}{3}k_B^2TN(\varepsilon_F),
\end{equation}
where $N(\varepsilon_F)$ is the DOS, $\varepsilon_F$ is the Fermi energy, and $k_B$ is the Boltzman constant. Considering the Mott relationship, thermopower probes the specific heat per electron: $S = C_e/ne$, where the units are V K$^{-1}$ for $S$, J K$^{-1}$ m$^{-3}$ for $C_e$, and m$^{-3}$ for $n$, respectively. However, it is common to express $\gamma = C_e/T$ in J K$^{-2}$ mol$^{-1}$ units. In order to focus on the $S/C_e$ ratio, let us define the dimensionless quantity,
\begin{equation}
q=\frac{S}{T}\frac{N_Ae}{\gamma},
\end{equation}
where $N_A$ is the Avogadro number, gives the number of carriers per formula unit (proportional to $1/n$).\cite{Behnia} The obtained $q$ = 0.90(3) is close to unity, suggesting about one hole per formula unit within the Boltzmann framework.\cite{Behnia}

Figure \ref{RMH}(a) show the effective field dependence of magnetization at various temperatures between 20 and 300 K for $\mu_0H//c$. Here $\mu_0H_{eff} = \mu_0(H-N_dM)$, where $N_d$ = 0.8 is the demagnetization factor. When $T < T_c$, the shape of $M(\mu_0H_{eff})$ curves is typical for ferromagnets, i.e., a rapid increase at low field with a saturation in higher magnetic field. The saturation magnetization $M_s$ decreases with increasing temperature, in line with the $M(T)$ curve.\cite{YuLIU} When $T > T_c$, it gradually changes into linear-in-field paramagnetic dependence at 300 K. Hall resistivity $\rho_{xy}(B)$ as a function of magnetic induction $B$ for tr-Cr$_5$Te$_8$ at the corresponding temperatures are depicted in Fig. \ref{RMH}(b). Here $B = \mu_0(H_{eff}+M) = \mu_0[H + (1-N_d)M]$. When $T < T_c$, the $\rho_{xy}(B)$ increases quickly at low $B$ region. With increasing $B$, the $\rho_{xy}(B)$ curve changes slightly with almost linear $B$ dependence at high $B$ region, similar to the shape of $M(\mu_0H_{eff})$ curve, indicating an AHE in tr-Cr$_5$Te$_8$.

In general, the Hall resistivity $\rho_{xy}$ in the ferromagnets is made up of two parts,\cite{Wang, Yan, WangY, Onoda2008}
\begin{equation}
\rho_{xy} = \rho_{xy}^O + \rho_{xy}^A = R_0B + R_s\mu_0M,
\end{equation}
where $\rho_{xy}^O$ and $\rho_{xy}^A$ are the ordinary and anomalous Hall resistivity, respectively. $R_0$ is the ordinary Hall coefficient from which apparent carrier concentration and type can be determined ($R_0 = 1/n_aq$), and $R_s$ is the anomalous Hall coefficient. With a linear fit of $\rho_{xy}(B)$ at high field region, the slope and intercept corresponds to $R_0$ and $\rho_{xy}^A$, respectively. Figure \ref{parameters}(a) presents the temperature dependence of $R_0$ and the derived $n_a$. The value of $R_0$ is positive, in line with the positive $S(T)$, confirming the hole-type carries. The derived carrier concentration $n_a$ increases abruptly around $T_c$ and decreases below about 180 K due to possible influence of spin reorientation on the Fermi surface. Note that Seebeck coefficient [Fig. 2(b)] shows similar temperature dependence suggesting its close connection with carrier concentration change, i.e. dominant diffusive mechanism. Given a weak temperature-dependent resistivity of 2.0 $\sim$ 2.8 $\mu\Omega$ m between 100 and 200 K [Fig. \ref{RSCT}(a)], the estimated carrier concentration $n_a\sim1.5\times10^{22}$ cm$^{-3}$ points to a mean free path $\lambda \sim$ 0.80(1) nm, comparable to the lattice parameters and close to the Mott-Ioffe-Regel limit.\cite{GunnarsonO} This is in agreement with its bad metal behavior. The carrier concentration $n_a\sim0.63\times10^{22}$ cm$^{-3}$ at 20 K corresponds to about 2 holes per formula unit, comparable with the estimation from $q$. On the other hand, the value of $R_s$ can be obtained by using $\rho_{xy}^A = R_s \mu_0 M_s$ with the $M_s$ taken from the linear fit of $M(\mu_0H_{eff})$ curves at high field region, which decreases monotonically with decreasing temperature and approaches almost zero at low temperature [Fig. \ref{parameters}(b)]. The value of $R_s$ is about two orders of magnitude larger than that of $R_0$.

The anomalous Hall conductivity $\sigma_{xy}^A$ ($\approx$ $\rho_{xy}^A / \rho_{xx}^2$) is presented in Fig. \ref{parameters}(c). Theoretically, the intrinsic contribution of $\sigma_{xy,in}^A$ is of the order of $e^2/(ha)$, where $e$ is the electronic charge, $h$ is the Plank constant, and $a$ is the lattice parameter.\cite{Onoda2006} Taking $a = V^{1/3} \sim 8.6$ {\AA} approximately, the $\sigma_{xy,in}^A$ is about 450 $\Omega^{-1}$ cm$^{-1}$. The calculated $\sigma_{xy}^A$ is much smaller than this value [Fig. \ref{parameters}(c)], which precludes the possibility of intrinsic mechanism. The extrinsic side-jump contribution of $\sigma_{xy,sj}^A$ is of the order of $e^2/(ha)(\varepsilon_{SO}/E_F)$, where $\varepsilon_{SO}$ and $E_F$ is the spin-orbital interaction energy and Fermi energy, respectively.\cite{Nozieres} The $\varepsilon_{SO}/E_F$ is usually less than $10^{-2}$ for the metallic ferromagnets. The side-jump mechanism, where the potential field induced by impurities contributes to the anomalous group velocity, follows a scaling behavior of $\rho_{xy}^A = \beta\rho_{xx}^2$, the same with the intrinsic mechanism. Figure \ref{parameters}(d) exhibits a clear linear relationship between $\rho_{xy}^A$ and $\rho_{xx}$ for tr-Cr$_5$Te$_8$ below $T_c$, further precluding the side-jump mechanism. This points to the possible skew-scattering mechanism which describes asymmetric scattering induced by impurity or defect could contribute to the AHE with scaling behavior of $\rho_{xy}^A = \beta\rho_{xx}$.

\section{CONCLUSIONS}

In summary, we investigated the transport properties and the AHE in tr-Cr$_5$Te$_8$ single crystals. The linear relationship between $\rho_{xy}^A$ and $\rho_{xx}$ reveals that the AHE in tr-Cr$_5$Te$_8$ is dominated by the extrinsic skew-scattering mechanism rather than the intrinsic mechanism or the extrinsic side-jump which gives the quadratic relationship between $\rho_{xy}^A$ and $\rho_{xx}$. With the rapid development of 2D materials for spintronics, further investigation of AHE in the nano-sheet of tr-Cr$_5$Te$_8$ is of interest.

\section*{Acknowledgements}
Work at Brookhaven is supported by the Research supported by the U.S. Department of Energy, Office of Basic Energy Sciences as part of the Computation Material Science Program (Y.L. and C.P.) and by the US DOE under Contract No. DE-SC0012704 (C.P.).


\begin{references}

\bibitem{Nagaosa} N. Nagaosa, J. Sinova, S. Onoda, A. H. MacDonald, and N. P. Ong, Phys. Mod. Phys. \textbf{82}, 1539 (2010).
\bibitem{Fang} Z. Fang, N. Nagaosa, K. S. Takahashi, A. Asamitsu, R. Mathieu, T. Ogasawara, H. Yamada, M. Kawasaki, Y. Tokura, and K. Terakura, Science \textbf{302}, 92 (2003).
\bibitem{Haldane} F. D. M. Haldane, Phys. Rev. Lett. \textbf{93}, 206602 (2004).
\bibitem{Xiao} D. Xiao, M. C. Chang, and Q. Niu, Rev. Mod. Phys. \textbf{82}, 1959 (2010).
\bibitem{Nakatsuji} S. Nakatsuji, N. Kiyohara, and T. Higo, Nature \textbf{527}, 212 (2015).
\bibitem{Takatsu} H. Takatsu, S. Yonezawa, S. Fujimoto, and Y. Maeno, Phys. Rev. Lett. \textbf{105}, 137201 (2010).
\bibitem{Shiomi} Y. Shiomi, M. Mochizuki, Y. Kaneko, and Y. Tokura, Phys. Rev. Lett. \textbf{108}, 056601 (2012).
\bibitem{Machida} Y. Machida, S. Nakatsuji, Y. Maeno, T. Tayama, T. Sakakibara, and S. Onoda, Phys. Rev. Lett. \textbf{98}, 057203 (2007).
\bibitem{Taguchi} Y. Taguchi, Y. Oohara, H. Yoshizawa, N. Nagaosa, and Y. Tokura, Science \textbf{291}, 2573 (2001).
\bibitem{Kubler} J. K\"{u}bler and C. Felser, Europhys. Lett. \textbf{108}, 67001 (2014).
\bibitem{Kiyohara} N. Kiyohara, T. Tomita, and S. Nakatsuji, Phys. Rev. Appl. \textbf{5}, 064009 (2016).
\bibitem{Nayak} A. K. Nayak, J. E. Fischer, Y. Sun, B. Yan, J. Karel, A. C. Komarek, C. Shekhar, N. Kumar, W. Schnelle, J. K\"{u}bler, C. Felser, and S. S. P. Parkin, Sci. Adv. \textbf{2}, e1501870 (2016).
\bibitem{Ueland} B. G. Ueland, C. F. Miclea, Y. Kato, O. A. Valenzuela, R. D. McDonald, R. Okazaki, P. H. Tobash, M. A. Torrez, F. Ronning, R. Movshovich, Z. Fisk, E. D. Bauer, I. Martin, and J. D. Thompson, Nat. Commun. \textbf{3}, 1067 (2012).
\bibitem{Surgers} C. S\"{u}rgers, G. Fischer, P. Winkel, and H. v. L\"{o}hneysen, Nat. Commun. \textbf{5}, 3400 (2014).
\bibitem{Suzuki} T. Suzuki, R. Chisnell, A. Devarakonda, Y. T. Liu, W. Feng, D. Xiao, J. W. Lynn, and J. G. Checkelsky, Nat. Phys. \textbf{12}, 1119 (2016).
\bibitem{Herbert} H. Ipser, K. L. Komarek, and K. O. Klepp, J. Less-Common Met., \textbf{92}, 265 (1983).
\bibitem{Street} G. B. Street, E. Sawatzky, and K. Lee, J. Phys. Chem. Solids, \textbf{34}, 1453 (1973).
\bibitem{Hamasaki} T. Hamasaki, and T. Hashimoto, Solid State Commun., \textbf{16}, 895 (1975).
\bibitem{Akram} M. Akram, and F. M. Nazar, J. Mater. Sci., \textbf{18}, 423 (1983).
\bibitem{Lukoschus} K. Lukoschus, S. Kraschinski, C. N\"{a}ther, W. Bensch, and R. K. Kremer, J. Solid State Chem., \textbf{177}, 951 (2004).
\bibitem{Huang1} Z. L. Huang, W. Bensch, S. Mankovsky, S. Polesya, H. Ebert, and R. K. Kremer, J. Solid State Chem., \textbf{179}, 2067 (2006).
\bibitem{Huang2} Z. L. Huang, W. Kockelmann, M. Telling, and W. Bensch, Solid State Sci., \textbf{10}, 1099 (2008).
\bibitem{Andresen} A. F. Andresen, Acta Chem. Scand., \textbf{24}, 3495 (1970).
\bibitem{Shimada} K. Shimada, T. Saitoh, H. Namatame, A. Fujimori, S. Ishida, S. Asano, M. Matoba, and S. Anzai, Phys. Rev. B, \textbf{53}, 7673 (1996).
\bibitem{YuLIU} Y. Liu and C. Petrovic, Phys. Rev. B \textbf{96}, 134410 (2017).
\bibitem{XiaoZ} X. Zhang, T. Yu, Q. Xue, M. Lei, and R. Jiao, J. Alloys Compd. \textbf{750}, 798 (2018)
\bibitem{XiaoH} X. Luo, W. Ren, and Z. Zhang, J. Magn. Magn. Mater. \textbf{445}, 37 (2018).
\bibitem{Ueda1} K. Ueda, and T. Moriya, J. Phys. Soc. Jpn., \textbf{39}, 605 (1975).
\bibitem{Liu} S. H. Liu, J. Appl. Phys. \textbf{35}, 1087 (1964).
\bibitem{Barnard} R. D. Barnard, \textit{Thermoelectricity in Metals and Alloys} (Taylor \& Francis, London, 1972).
\bibitem{Behnia} K. Behnia, D. Jaccard and J. Flouquet, J. Phys.: Condens. Matter. \textbf{16}, 5187 (2004).
\bibitem{Wang} Q. Wang, S. S. Sun, X. Zhang, F. Pang, and H. C. Lei, Phys. Rev. B \textbf{94}, 075135 (2016).
\bibitem{Yan} J. Yan, X. Luo, F. C. Chen, Q. L. Pei, G. T. Lin, Y. Y. Yan, L. Hu, P. Tong, W. H. Song, X. B. Zhu, and Y. P. Sun, Appl. Phys. Lett. \textbf{111}, 022401 (2017).
\bibitem{WangY} Y. H. Wang, C. Xian, J. Wang, B. J. Liu, L. S. Ling, L. Zhang, L. Cao, Z. Qu, and Y. M. Xiong, Phys. Rev. B \textbf{96}, 134428 (2017).
\bibitem{Onoda2008} S. Onoda, N. Sugimoto, and N. Nagaosa, Phys. Rev. B \textbf{77}, 165103 (2008).
\bibitem{GunnarsonO} O. Gunnarson, M. Calandra and J. E. Han, Rev. Mod. Phys. \textbf{75}, 1085 (2003).
\bibitem{Onoda2006} S. Onoda, N. Sugimoto, and N. Nagaosa, Phys. Rev. Lett. \textbf{97}, 126602 (2006).
\bibitem{Nozieres} P. Nozi\`{e}res and C. Lewiner, J. Phys. (Paris) \textbf{34}, 901 (1973).

\end{references}
\end{document}